\documentclass[12pt,letterpaper]{article}
\usepackage{amssymb}
\usepackage{amsmath}
\usepackage{amscd}
\usepackage{graphicx}
\usepackage{tikz}
\usepackage{subfigure}
\usepackage{array}
\usepackage{pdflscape}
\usepackage{tensor}
\usepackage{ulem}
\usepackage{slashed}
\def\be{\begin{equation}}
\def\ee{\end{equation}}
\def\bea{\begin{eqnarray}}
\def\eea{\end{eqnarray}}
\def\bse{\begin{subequations}}
\def\ese{\end{subequations}}

\def\bna{\bql\begin{array}{rcl}}
\def\ena{\end{array}\eql}
\def\bnn{\beq\begin{array}{rcl}}
\def\enn{\end{array}\ee}
\def\bet{\begin{tabular}}
\def\bsf{\sffamily\bfseries}

\def\4{\text{\bsf-}}

\definecolor{Red}    {rgb}{1.00,0.00,0.00} 
\definecolor{Green}  {rgb}{0.00,0.75,0.00} 
\definecolor{Blue}   {rgb}{0.00,0.00,1.00} 
\definecolor{Orange} {rgb}{1.00,0.67,0.00} 
\definecolor{Purple} {rgb}{0.50,0.00,0.50} 
\definecolor{Gold}   {rgb}{1.00,0.90,0.00} 
\definecolor{Magenta}{rgb}{1.00,0.00,1.00} 
\definecolor{Turque} {rgb}{0.00,0.90,0.90} 
\definecolor{Seaweed}{rgb}{0.00,0.25,0.00} 
\definecolor{Brown}  {rgb}{0.50,0.13,0.00} 
\definecolor{Cobalt} {rgb}{0.00,0.00,0.50} 
\definecolor{Sage}   {rgb}{0.00,0.50,0.38} 
\definecolor{grey1}  {rgb}{0.20,0.20,0.20} 
\definecolor{grey2}  {rgb}{0.40,0.40,0.40} 
\definecolor{grey3}  {rgb}{0.60,0.60,0.60} 
\definecolor{grey4}  {rgb}{0.80,0.80,0.80} 
\definecolor{grey5}  {rgb}{0.90,0.90,0.90} 
\def\C#1#2{{\ifcase#1\or
             \color{Red}\or\color{Green}\or\color{Blue}\or
              \color{Orange}\or\color{Purple}\or\color{Gold}\or
             \color{Magenta}\or\color{Turque}\or\color{Seaweed}\or
               \color{Brown}\or\color{Cobalt}\or\color{Sage}\or
                 \color{grey1}\or\color{grey2}\or\color{grey3}\or
                 \color{grey4}\else\color{grey5}\fi#2}}
\definecolor{gray}{rgb}{.7,.7,.7}
\def\XXX{\colorbox{yellow}{\color{red}\bf X\kern-4pt{\Large$\bs*$}\kern-4.125ptX}}
\def\[{\left[}
\def\]{\right]}

\font\ro=cmsy10                          
\def\kcr{{\hbox{\ro \char'170}}}                
\def\ktl{{\hbox{\ro \char'170}}}        
\def\ktr{{\hbox{\ro \char'170}}}        
\def\kbl{{\hbox{\ro \char'170}}}        
\def\kbr{{\hbox{\ro \char'170}}}        

\parskip=\medskipamount                 
\thicklines                         

\newskip\humongous \humongous=0pt plus 1000pt minus 1000pt

\newif\ifdtup

\def\border{                                            
        \setlength{\unitlength}{1mm}
        \newcount\xco
        \newcount\yco
        \xco=-21
        \yco=12
        \begin{picture}(140,0)
        \put(\xco,\yco){$\ktl$}
        \advance\yco by-1
        {\loop
        \put(\xco,\yco){$\kcr$}
        \advance\yco by-2
        \ifnum\yco>-240
        \repeat
        \put(\xco,\yco){$\kbl$}}
        \xco=158
        \yco=12
        \put(\xco,\yco){$\ktr$}
        \advance\yco by-1
        {\loop
        \put(\xco,\yco){$\kcr$}
        \advance\yco by-2
        \ifnum\yco>-240
        \repeat
        \put(\xco,\yco){$\kbr$}}
        \put(-19.75,13){\tiny **University of Maryland * Center for String and
         Particle  Theory * Physics Department**University of Maryland * Center
        for String and Particle  Theory** }
        \put(-19.75,-241.5){\tiny **University of Maryland * Center for String and
         Particle  Theory * Physics Department**University of Maryland * Center
        for String and Particle  Theory** }
        \end{picture}
        \par\vskip-8mm}

\def\headpic{                                           
        \indent
        \setlength{\unitlength}{.4mm}
        \thinlines
        \par
        \begin{picture}(29,16)
        \put(165,16){\line(1,0){4}}
        \put(170,16){\line(1,0){4}}
        \put(180,16){\line(1,0){4}}
        \put(175,0){\line(1,0){4}}
        \put(180,0){\line(1,0){4}}
        \put(185,0){\line(1,0){4}}
        \put(169,0){\line(0,1){16}}
        \put(170,0){\line(0,1){16}}
        \put(179,0){\line(0,1){16}}
        \put(180,0){\line(0,1){16}}
        \put(184,0){\line(0,1){16}}
        \put(185,0){\line(0,1){16}}
        \put(169,16){\oval(8,32)[bl]}
        \put(170,16){\oval(8,32)[br]}
        \put(179,0){\oval(8,32)[tl]}
        \put(185,0){\oval(8,32)[tr]}
        \end{picture}
        \par\vskip-6.5mm
        \thicklines}
\def\endtitle{\end{quotation}\newpage}                  

\begin{document}

\border\headpic {\hbox to\hsize{\today \hfill
{PP 012-017}}}
\par \noindent
{ \hfill
{math-ph/1306.0550}}
\par

\par

\setlength{\oddsidemargin}{0.5in}
\setlength{\evensidemargin}{-0.5in}
\begin{center}
\vglue .10in
{\large\bf Creating Infinitesimal Generators And Robust Messages With Adinkras}\\[.8in]

Keith\, Burghardt\footnote{keith@umd.edu}
\\[1.5in]

{\it Center for String and Particle Theory\\
Department of Physics, University of Maryland\\
College Park, MD 20742-4111 USA}\\[1.1in]

{\bf ABSTRACT
}
\\[.01in]
\end{center}
\begin{quotation}
{
Adinkras are graphs that can describe off-shell supermultiplets in 1 dimension with a Lie superalgebra known as Garden algebra. In this paper, I show that the degrees of freedom of the adinkra can be represented by a subgraph called a baobab. Because the structure of adinkras and baobabs are very general, I will show that all finite-dimensional Lie superalgebras can be similarly described by more general Lie adinkras and Lie baobabs. Furthermore, it will be shown that adinkras can represent forward error correction block codes, and bit erasures in Garden algebra adinkras can be corrected using logic circuits derived from baobabs.}

\endtitle

\begin{center}
${~~~}$ \newline
``$\text{\it{\small We think in generalities, but we live in details.}}$"${~~~}$
\end{center}
$~~~~~~~~~~~~~~~~~~~~~~$ {\small -- Alfred North Whitehead}
\newline ${~~~}$
\noindent

\section{Introduction}

$~~\ ~$ Adinkras are graphs of 1 dimensional off-shell supermultiplets\footnote{For a thorough introduction to adinkras, see \cite{Adinkra} or \cite{YanThesis}.}. Each superdifferential operator in the supermultiplet, $D_I$, is represented as an edge with some color $I$, while bosons and fermions are represented by open and closed nodes respectively (see Fig. \# 1). Edges connect open and closed nodes together to represent the transform of a fermion to a boson, or vice versa, by the super-differential operator $D_I$. When a node, representing a field $\eta$ is directed to another node (with field $\eta'$) by an edge with the color $I$ then $D_I \eta = \alpha' \eta'$ and $D_I \eta' = \alpha \frac{d}{d\tau}\eta$ where $\alpha, \alpha' \in \mathbb{C}$. Furthermore, when a dashed edge with color $I$ forms an arc between  $\eta$ and $\eta'$ then the transformation is the same as before, except $D_I \eta$ also multiplies $\eta'$ by $-1$ and vice versa.

The relation between super-differential operators,
\be
\{D_I,D_J\}=2\hskip0.012in\dot{\imath}\hskip0.012in\frac{d}{d\tau}\hskip0.012in\delta_{ij}
\ee 
 defines Garden algebra, and implies that an odd number of dashed edges are around every 2-color cycle to represent $D_I D_J = -D_J D_I$ when $I\neq J$. Furthermore, to satisfy (1), whenever a path reaches a closed node, the representative transformation picks up a phase, $\dot{\imath}$. Lastly, this algebra non-trivially implies that adinkra topologies are hypercubes quotiented by ``doubly even" codes whose weight is $0 (mod~ 4)$ (and vice versa) [7, 14, 15].
 
An example of two adinkras are in Fig. \# 1. In Fig. \# 1 (a), labeling the fermion nodes $\Psi_i$ and the boson nodes $\Phi_j$,  the representative transformations are:
\newline\newline
$~~~~~~~~~~~~~~~~~~~~~~~~~~~~~~{\color{red}D_{red}}\Psi_1= \Phi_2~~~~~~~~~~~~~~~~~{\color{red}D_{red}}\Psi_2= -\Phi_1$\newline
$~~~~~~~~~~~~~~~~~~~~~~~~~~~~~~{\color{red}D_{red}}\Phi_2=\dot{\imath} \frac{d}{d\tau}\Psi_1~~~~~~~~~~~~~{\color{red}D_{red}}\Phi_1=-\dot{\imath}\frac{d}{d\tau} \Psi_2$\newline
$~~~~~~~~~~~~~~~~~~~~~~~~~~~~~~{\color{blue}D_{blue}}\Psi_1= \Phi_1~~~~~~~~~~~~~~~~{\color{blue}D_{blue}}\Psi_2=\Phi_2$\newline
$~~~~~~~~~~~~~~~~~~~~~~~~~~~~~~{\color{blue}D_{blue}}\Phi_2=\dot{\imath}\frac{d}{d\tau} \Psi_2~~~~~~~~~~~~{\color{blue}D_{blue}}\Phi_1=\dot{\imath}\frac{d}{d\tau} \Psi_1~~~~~~~~~~~~~~~~~~~~~~~(2)$\newline\newline
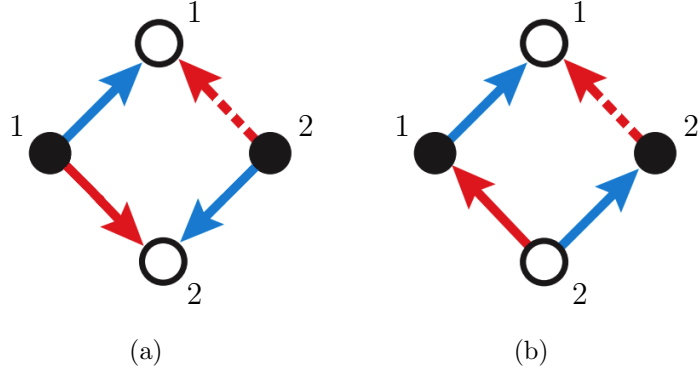
\begin{figure}[!htbp]
\centering
\subfigure[]{\label{f:diamondPath}
\includegraphics[width=0.3\columnwidth]{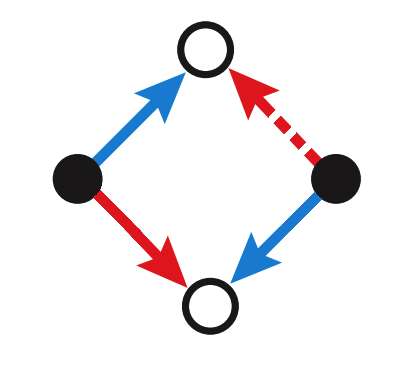}}
\put(-47,3){$2$}      \put(-47,97){$1$}
\put(-106,59){$1$}      \put(-10,59){$2$}
\quad
\subfigure[]{\label{f:bowtiePath}
\includegraphics[width=0.3\columnwidth]{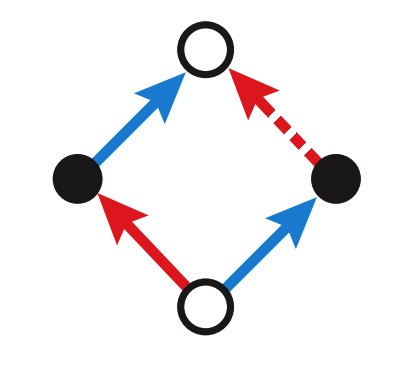}} 
\put(-47,3){$2$}         \put(-47,97){$1$}
\put(-106,59){$1$}      \put(-10,59){$2$}
\label{f:diambtPath}
\caption{Two-color adinkras with labels for bosons (open nodes) and fermions (closed nodes).}
\end{figure}

In comparison, the representative transformations for Fig. \# 1(b) are:
\newline\newline
$~~~~~~~~~~~~~~~~~~~~~~~~~~~~~~{\color{red}D_{red}}\Psi_1=\frac{d}{d\tau} \Phi_2~~~~~~~~~~~~~~{\color{red}D_{red}}\Psi_2=- \Phi_1$\newline
$~~~~~~~~~~~~~~~~~~~~~~~~~~~~~~{\color{red}D_{red}}\Phi_2=\dot{\imath} \Psi_1~~~~~~~~~~~~~~~~{\color{red}D_{red}}\Phi_1=-\dot{\imath} \frac{d}{d\tau} \Psi_2$\newline
$~~~~~~~~~~~~~~~~~~~~~~~~~~~~~~{\color{blue}D_{blue}}\Psi_1= \Phi_1~~~~~~~~~~~~~~~~{\color{blue}D_{blue}}\Psi_2=\frac{d}{d\tau} \Phi_2$\newline
$~~~~~~~~~~~~~~~~~~~~~~~~~~~~~~{\color{blue}D_{blue}}\Phi_2=\dot{\imath} \Psi_2~~~~~~~~~~~~~~~{\color{blue}D_{blue}}\Phi_1=\dot{\imath}\frac{d}{d\tau} \Psi_1~~~~~~~~~~~~~~~~~~~~~~~(3)$
\newline\newline
 These graphs represent several superfluous equations, however, which obscure the differences between the supermultiplets. For example, in the above equations, ${\color{red}D_{red}}{\color{blue}D_{blue}} \Psi_1=\dot{\imath}\frac{d}{d\tau}\Psi_2 \implies {\color{blue}D_{blue}}{\color{red}D_{red}} \Psi_1=-\dot{\imath}\frac{d}{d\tau}\Psi_2$ because $D_I D_J = -D_J D_I$. This is the motivation behind a new graph introduced in this paper, called a baobab (or Garden baobab when we specifically mean baobabs for Garden algebra adinkras), which can distinguish supermultiplets with significantly fewer edges than an adinkra.

\begin{figure}[!htbp]
  \centering
   \subfigure[]{\label{f:bowtiemin}
   \includegraphics[width=0.3\columnwidth]{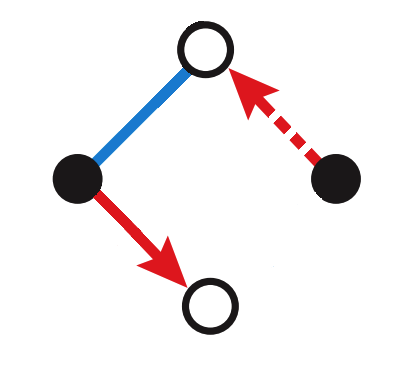}}
   \quad
   \subfigure[]{\label{f:diamondmin}
   \includegraphics[width=0.3\columnwidth]{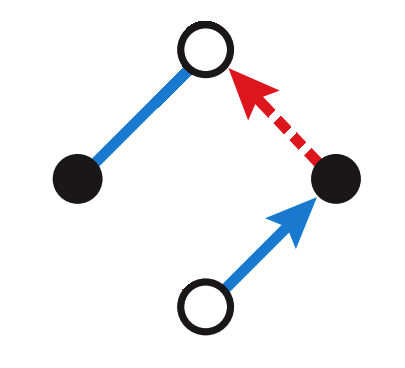}} 
   \label{f:diambtMin}
   \caption{An example of a Garden baobab for each adinkra in Fig. \# 1.}
\end{figure}

The Garden baobab is a tree-like mixed graph that is defined to represent all the free parameters necessary to reconstruct adinkras using Garden algebra relations. Garden baobabs usually have few or no cycles, and only about half (and sometimes fewer) of their edges are directed. Adinkras usually require large matrices to determine isomorphism classes, even though a large majority of the non-zero elements are unnecessary. Baobabs, however, require a minimal number of elements, possibly allowing for the isomorphism class to be quickly determined with a modified version of the previous isomorphism algorithms [4, 13]. 

As an example of a baobab, the adinkras in Fig. \# 1 can be reconstructed from the Garden baobabs in Fig. \# 2.

Note, however, that the overall structure of an adinkra doesn't seem to be unique to Garden algebra. This sets up an enticing question: can other Lie superalgebras be organized into an adinkra-like graph? 

Previous work has found that representations of the Clifford algebra, C$\ell_{0,m}$,
could be graphically represented by adinkras, but this was a fairly special case \cite{Clifford}. In addition, there was some work relating adinkras to the exterior algebra, but this was, again, a special case, whose representation was not fully explored \cite{Naples}. I will show in this paper that all finite dimensional Lie superalgebras (or more precisely any Lie superalgebra whose elements have a real representation) can be described by Lie adinkras. To demonstrate the new flexibility, we can compactly describe $\mathbb{H}$, a sub-algebra of C$\ell_{0,3}$, using a small set of matrices.
 \vskip0.2in
$
{\color{blue}i}={\color{blue}\left[ \begin{array}{cccc}
0& 1&0&0\\ 
-1 & 0&0&0\\
0&0&0&-1\\
 0&0&1&0
\end{array} \right]}~~~~~
{\color{Green}j}={\color{Green}\left[ \begin{array}{cccc}
0&0&1&0\\ 
0 & 0&0&1\\
-1&0&0&0\\
 0&-1&0&0
\end{array} \right]}~~~~~
{\color{red}k}={\color{red}\left[ \begin{array}{cccc}
0&0&0&1\\ 
0 & 0&-1&0\\
0&1&0&0\\
 -1&0&0&0
\end{array} \right]}$
\vskip1.7in
  \begin{picture}(-20,0)
 \put(40,00){  \includegraphics[width=0.3\columnwidth]{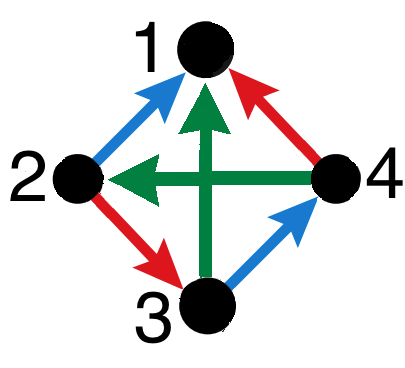}}
 \put(200,0){  \includegraphics[width=0.3\columnwidth]{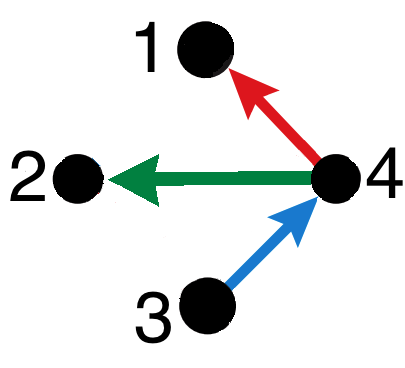}} 
  \end{picture}
\vskip0.0in
\noindent
Figure 3: The real representation of the quaternion elements and their graphical depiction (left). The right diagram depicts the free parameters of the adinkra. 
 \newline
 
 Let the blue, green, and red edges in Fig, \# 3 represent their respective matrices, then the quaternion adinkra graphically depicts the relations between the quaternion elements ${\color{blue}i}$, ${\color{Green}j}$, and ${\color{red}k}$. Although it was previously shown that Clifford algebra representations can be constructed from adinkras, C$\ell_{0,3}$ is an $8\times 8$ matrix represented by a 3-cube adinkra. The new adinkra creates matrices that are dimensionally reduced, and graphically simpler than the regular adinkra to represent quaternions.

 Edges directed from node $\bullet_m\leftarrow  \bullet_n$ imply the representation has the element $a^m_{n}=-1$, and $a^n_{m}=1$.  The free parameters of the adinkra can be represented by the graph on the lower right. Using the property that ${\color{blue} i} {\color{Green} j } {\color{red} k} = -1$ and $\{{\color{blue}i},{\color{Green}j}\}=\{{\color{blue}i},{\color{red}k}\}=\{{\color{Green}j},{\color{red}k}\}=0$, one can reconstruct the ``quaternion" adinkra from the quaternion baobab, as seen in Fig. \# 4. The crossed out diagram has the relation:

$
~~~~~~~~~~~~~~~~~~~~~~~~~~~~~{\color{blue} i} {\color{Green} j } {\color{red} k} = \left[ \begin{array}{cccc}
-1& 0&0&0\\ 
0 & 1&0&0\\
0&0&-1&0\\
 0&0&0&1
\end{array} \right]~~~~~~~~~~~~~~~~~~~~~~~~~~~~~~~~~~~~~~~~(4)$  
\vskip0.2in
\noindent
which disagrees with quaternion algebra. Therefore, there is exactly one possible adinkra that can be constructed from the baobab using the Lie bracket values.
  
  \vskip4.41in
  \begin{picture}(-20,0)
 \put(40,20){  \includegraphics[width=0.8\columnwidth]{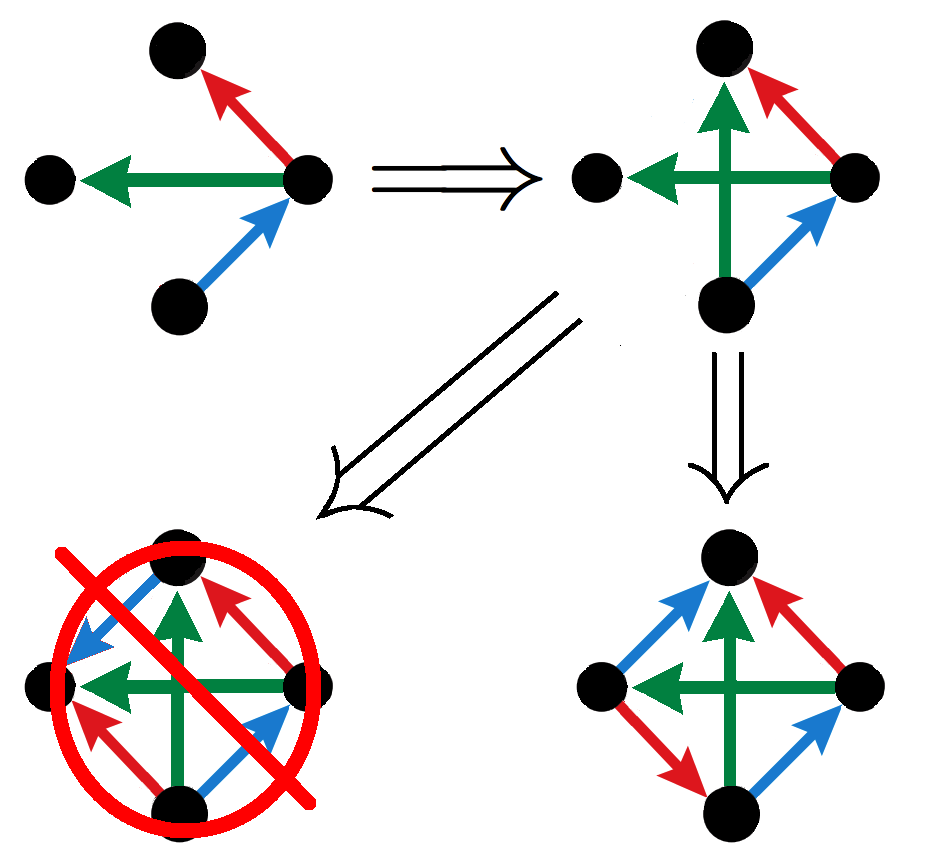}}
    \end{picture}
\noindent
Figure 4: Building the adinkra for the quaternion algebra from one of its associated baobabs. \vskip0.2in

 Lie adinkras allow for relationships between representations to be immediately determined,  while Lie baobabs generalize Garden baobabs in order to more easily distinguish non-isomorphic representations of the same algebra. When two Lie adinkras represent distinct algebras, they can be compared and contrasted alike to Dynkin diagrams, although they are unique in being able to re-create and compare representations instead of the just the algebra.
   
Interestingly, Lie adinkras and baobabs seem to have a deep connection with information theory. It was previously shown that one can determine the ``Garden" adinkra's {\it chromotopology} (topology, edge colors, and colored nodes) by quotienting the graph by a doubly even code\footnote{See  \cite{Clifford},  \cite{Code}, or  \cite{Chrome}, for example, for explanations of Garden adinkra chromotopology. Furthermore, \cite{Code}, \cite{Code2}, \cite{Code3} and \cite{YanThesis} are all good sources to understand doubly even codes and their relation to Garden algebra.}. In this paper, I show that one can determine chromotopology, edge dashing, and edge directions of Garden adinkras, quaternion adinkras, and possibly others, using a few starting ``input" bits to a series of gates. If dashed edges in a Garden adinkra are represented as a ``0", and undashed edges, ``1", then the gate $\neg(x\oplus y\oplus z)$, for $x,~y,~z \in \mathbb{Z}_2$, which I call the ``Not Double Exclusive Or" (NDXOR) gate, determines the fourth edge of a two-color loop given three edge dashings $x$, $y$, and $z$. A similar gate determines the directed edges of a Garden adinkra given the directed edges of a Garden baobab. These gates can be set up in ``cascade" where the output of one gate becomes the input of another via the baobab topology, allowing for the weights of all edges in a Garden adinkra to be re-created.

Furthermore, these graphs seem to represent a new set of forward error correction block codes, because many edges, representing bits of a message, could be changed in a Lie adinkra and still detected as long as at least one Lie baobab subgraph is correct.

The paper will be organized as follows: section 2 will introduce the representations encoded in Garden adinkras. These will be used in section 3 to determine the edge dashing and general structure of Garden baobabs. Section 4 will show how certain finite dimensional Lie superalgebras can be encoded into a Lie adinkra and Lie baobab. Section 5 explores the relationship between these graphs and information theory, while section 6 concludes the paper.

\vskip0.7in
\section{Supermultiplets as Garden Adinkras}

$~~~$Supermultiplets are representations of SUSY algebras, which transform a vector of bosons into a vector of fermions and vice versa. In one dimension, many of these supermultiplets follow Garden algebra. 
Let the matrices that permute fermions on the right hand side, be labeled ${\bf D_{I{}_L}}$ and matrices that permute bosons, ${\bf D_{I{}_R}}$. The operators $D_I D_J$ acting on bosons will therefore be \newline\newline
$
~~~~~~~~~~~~~~~~~~~~~~~~~~~~~~~~~~~~~~~~~D_I D_J {\bf \Phi} = {\bf D_{I{}_R}}{\bf D_{J{}_L}} {\bf \Phi}~~~~~~~~~~~~~~~~~~~~~~~~~~~~~~~~~~~~~~~~~~~~~(7)
$\newline\newline
\noindent in matrix form. 

Garden algebra now takes the form:
\newline\newline
$
 ~~~~~~~~~~~~~~~~~~~~~~~~~~~~~~~~~~~~{\bf D_{I{}_R}}{\bf D_{J{}_L}}+ {\bf D_{J{}_R}}{\bf D_{I{}_L}} = 2 \frac{d}{d\tau}{\bf I} \delta_{I J}
 $\newline
$~~~~~~~~~~~~~~~~~~~~~~~~~~~~~~~~~~~~~{\bf D_{I{}_L}}{\bf D_{J{}_R}}+ {\bf D_{J{}_L}}{\bf D_{I{}_R}} = 2 \frac{d}{d\tau}{\bf I} \delta_{I J}~~~~~~~~~~~~~~~~~~~~~~~~~~~~~~~~~~(8)$
\newline
\newline
To prove a proposition in the next section, it is worth noticing some of the properties of these graphs implied by (8):

\begin{itemize}

\item Ignoring derivatives, $\frac{d}{d\tau}$, ${\bf D_{I{}_R}}^T={\bf D_{I{}_L}}$ and ${\bf D_{I{}_R}}{\bf D_{J{}_L}}=-({\bf D_{I{}_R}}{\bf D_{J{}_L}})^T,~I\neq J$.

\item The dimensions of the matrices are $2^{n-1} \times 2^{n-1}$

\end{itemize}

\noindent
both of which come from observations made previously on similar R- and L- matrices encoded in a Garden adinkra [2, 7].
 
Finally note that the matrices:\newline\newline
$
~~~~~~~~~~~~~~~~~~~~~~~~~~~~~~~\left
\{\Gamma_I = 
\left( \begin{array}{cc}
{\bf 0} &{\bf D_{I{}_L}} \\ 
{\bf D_{I{}_R}} & {\bf 0}\\ 
\end{array} \right) \right\}
~\forall~ \text{\bf I} \in \{1,...,n+k\}~~~~~~~~~~~~~~~~~~~~~~~~~~~~~~~~(9)
$\vskip0.2in
\noindent
 represent the supermultiplet (here {\bf 0} is a $2^{n-1} \times 2^{n-1}$ block matrix of 0s), therefore a Garden adinkra is a graphical representation of the Garden algebra's elements. 

Equation (8) provides a set of linear equations that the matrix elements must satisfy.  The constraints the equations pose, however, imply that the dashing of one edge is dependent on the dashing of another, and similarly for directed edges. Therefore, a smaller graph must exist that represents the free parameters of a Garden adinkra: the Garden baobab.

\vskip0.7in
\section{The Free Parameters Of A Garden Adinkra}

$~~~$In this section, I will determine the minimum information needed to reproduce a Garden adinkra, which will be compactly represented by a Garden baobab. Firstly, to determine the edge dashing degrees of freedom, I will use the following proposition.\newline

\noindent
{\bf Proposition 1:} There are $2^n + (k-1)$ dashing degrees of freedom in a Garden adinkra with $n+k$ edge colors.
 
\noindent {\bf Proof:}\newline 
By ignoring derivatives in ${\bf D_{I{}_{R/L}}}$ and ${\bf D_{I{}_{R/L}}}$, (8) says that $({\bf D_{I{}_L}}{\bf D_{J{}_R}})^i_j=-({\bf D_{I{}_L}}{\bf D_{J{}_R}})^j_i$ (the two equations together in (8) are redundant if the matrices are square), which reduces the degrees of freedom between the elements by 1. Therefore, there are 3 degrees of freedom between the elements  $({\bf D_{I{}_L}}{\bf D_{J{}_R}})^i_j$ and  $({\bf D_{I{}_L}}{\bf D_{J{}_R}})^j_i$ which can be arbitrarily partitioned between ${\bf D_{I{}_L}}^i_k$, ${\bf D_{J{}_R}}^k_j$, ${\bf D_{I{}_L}}^j_\ell$, and ${\bf D_{J{}_R}}^\ell_i$. Because of this freedom, one could partition these edge colors such that there are $2^{n-1}$ degrees of freedom for ${\bf D_{I{}_R}}$, and $2^{n-2}$ degrees of freedom for ${\bf D_{J{}_R}}$. Along these lines, each additional edge color has 1/2 as many degrees of freedom as the one before it. For example, an edge color ${\bf D_{K{}_{R/L}}}$ must obey the two equations $({\bf D_{I{}_L}}{\bf D_{K{}_R}})^i_j=-({\bf D_{I{}_L}}{\bf D_{K{}_R}})^j_i$ and $({\bf D_{J{}_L}}{\bf D_{K{}_R}})^i_j=-({\bf D_{J{}_L}}{\bf D_{K{}_R}})^j_i$, therefore ${\bf D_{K{}_{R/L}}}$ has $(2^{n-1}/2)/2 = 2^{n-3}$ degrees of freedom.
Therefore, the dashing degrees of freedom for the edge colors in an $n$-cube Garden adinkra is $2^n + 2^{n-1}+2^{n-2}+...=2^n-1$. More edge colors add exactly one additional degree of freedom because ${\bf D_{M{}_L}}{\bf D_{P{}_R}}+ {\bf D_{P{}_R}}{\bf D_{M{}_L}} $ is the same  for $ {\bf D_{M{}_R}}\leftrightarrow -{\bf D_{M{}_R}}$. \begin{flushright}$\Box$\end{flushright}

This is equivalent to Thm. \# 3.5.4 in Dr. Yan Zhang's thesis, although it was found in a completely distinct fashion \cite{YanThesis}.

Similarly, one can put restrictions on the number of directed edge degrees of freedom. \newline

\noindent
{\bf Proposition 2:} There are no fewer than $n$ and no more than $2^{n-1}$ degrees of freedom for directed edges in a Garden adinkra.
 
\noindent {\bf Proof:}\newline 
If $n-1$ directed edges are known in an adinkra, then there exists at least one edge color whose edge directions are unknown. Trivially, if two edge colors, $I$ and $J$, have unknown edge directions, then every 2-color cycle with colors $(I,J)$ could be directed almost arbitrarily. If all but one edge color contains a directed edge, then there exists two colors such that every 2-color cycle contains at most one directed edge. It is easy to see that knowing the direction of one edge will not determine all other directed edges in the cycle. Therefore, an adinkra contains at least $n$ directed edge degrees of freedom

There are no more than $2^{n-1}$ degrees of freedom, on the other hand, because the edge directions of open nodes determine the edge directions of closed nodes and vice versa. The only way Garden adinkras can change their directed edges while preserving Garden algebra is by switching the head of each arc connected to a node with the tail and vice versa. At maximum $2^{n-1}$ nodes can do so.\begin{flushright}$\Box$\end{flushright}

With the previous propositions, one can now describe the Garden baobab graph. Proposition \# 3 will show that an $n$-color adinkra's dashing degrees of freedom can be described by a spanning tree, while proposition \# 4 will show how the dashing degrees of freedom of $n+k$-color adinkra can be described with $k$ cycles in addition to a spanning tree. \newline

\noindent
{\bf Proposition 3:} A Garden baobab with $n$ edge colors is a spanning tree.
 
\noindent {\bf Proof:}\newline  
Recall from Prop. \# 1, that given $2p$ degrees of freedom in one matrix, there are $p$ degrees of freedom in another matrix because of (8). It is equally valid for both matrices to share an arbitrary number of degrees of freedom, by applying the property to individual elements, $({\bf D_{P{}_R}})^i_j$. For this reason, degrees of freedom need not be grouped in a specific order. 

Also notice that any cycle in an $n$-color adinkra will contain each edge color an even (including 0) number of times\footnote{Hypercubes can have nodes represented as boolean vectors, with adjacent node's codewords differing by 1 \cite{Hypercube}. It has been shown that each edge color in an $n$-color (hypercube) adinkra can be represented as codewords with weight 1, such that a node $o$ connects to a node $p$ if $o\oplus D_I = p$ (mod 2) for some $D_I$ [5, 14, 15]. Therefore, in order for a cycle to reproduce the same boolean vector for each node, every edge color must appear an even number of times.}.

Because dashed edges can be directed, the Garden baobab needs only $2^n-1$ edges. If the graph contains a cycle $(I, J,..., L)$ then by permuting operators, $(D_I, D_J,..., D_L)$ is a known $\mathbb{C}$ number with some number of derivatives, implying cycles do not add additional information about the adinkra. One edge's information can be derived from other edges in the cycle, creating a redundancy. Because an $n$-cube spanning trees have $2^{n}-1$ edges, and because the baobab of an $n$-color adinkra is necessarily cycle-free, the baobab is a spanning tree.
\begin{flushright}$\Box$\end{flushright}

The converse is not necessarily true in general. For example, a Hamilton path from one end of a cube to another cannot determine whether an adinkra has exactly one source and sink, or more. The converse seems to be true, however, for all ``valise" adinkras with $2^{n-1}$ sources and sinks respectively (see Fig. \# 6).

\noindent
{\bf Proposition 4:} An adinkra with $n+k$ edge colors adds $k$ cycles to an underlying spanning tree in the baobab such that each cycle contains an edge color that is unique to that cycle.
 
\noindent {\bf Proof:}\newline 
With $k$ additional edge colors, a subgraph of a baobab is a spanning tree for the same reason as before. Additional cycles, however, must be added to a baobab to represent the $k$ additional dashing degrees of freedom.

 Let me define the odd edge color set, $\mathcal{O}_i$, of a cycle, $\mathcal{C}_i,~1\le i\le k$, to be all edge colors that appear in the cycle an odd number of times. 
 What I will prove is that $k$ cycles are independent of each other, and thus represent a distinct degree of freedom if and only if  $\exists~ D_M \in \mathcal{O}_i$,  such that $D_M\notin \mathcal{O}_j~\forall~j\neq i$.
 
Assume that $\forall~ i~ \exists~ D_M \in \mathcal{O}_i$,  such that $D_M\notin \mathcal{O}_j~j\neq i$. The edge dashing of one cycle clearly cannot determine the edge dashing of the other $k-1$ cycles, and furthermore, removing one edge from each cycle creates a spanning tree with $2^n-1$ degrees of freedom, therefore there are $k$ additional degrees of freedom. To prove the converse, assume $k$ cycles contain all the additional dashing degrees of freedom, but $\exists~ \mathcal{O}_\ell \subset \mathcal{O}_1\cup\text{...}\cup \mathcal{O}_k$. $\mathcal{C}_\ell$ cannot represent a new degree of freedom, because, any edge $D_I \in \mathcal{O}_\ell$ can be part of a cycle $\mathcal{C}_{j\neq \ell}$. In other words, the information contained in $\mathcal{O}_\ell$ can really be produced from the set of cycles $\mathcal{C}_j,~j\neq \ell$. 

 Therefore a Garden baobab with $k$ such cycles contains $2^n+(k-1)$ dashing degrees of freedom. \begin{flushright}$\Box$\end{flushright}

Using Prop. \# 3, the lower bound for the number of directed edges is tight, because Garden adinkras similar to Fig. \# 5, are determined by $n$ directed edges (and are easy to generalize).
Similarly, upper limit is found when looking at Garden adinkras similar to Fig. \# 6, which are called ``valise" adinkras  \cite{Clifford}. The heads and tails of arcs connected to any node could switch and still obey Garden algebra.
\newline
\vskip0.5in
  \begin{picture}(-20,0)
  \put(0,-50){ \includegraphics[width=0.45\columnwidth]{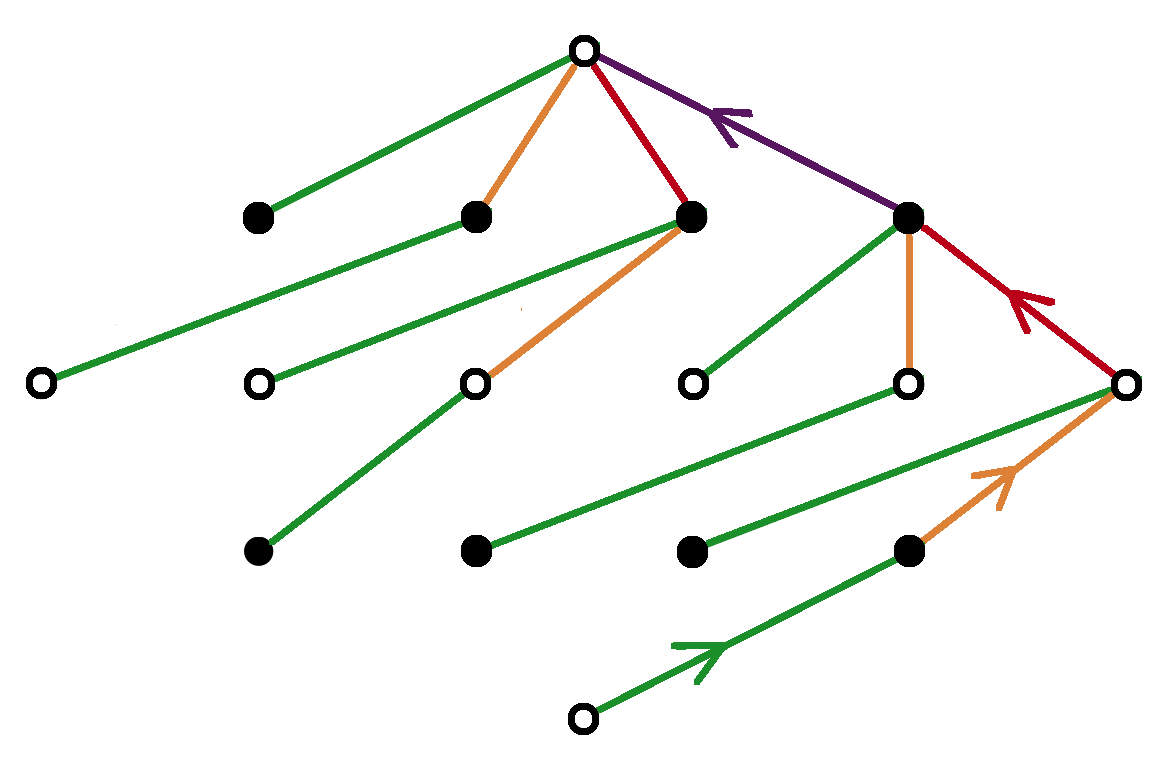}}
      \put(175,-50){ \includegraphics[width=0.42\columnwidth]{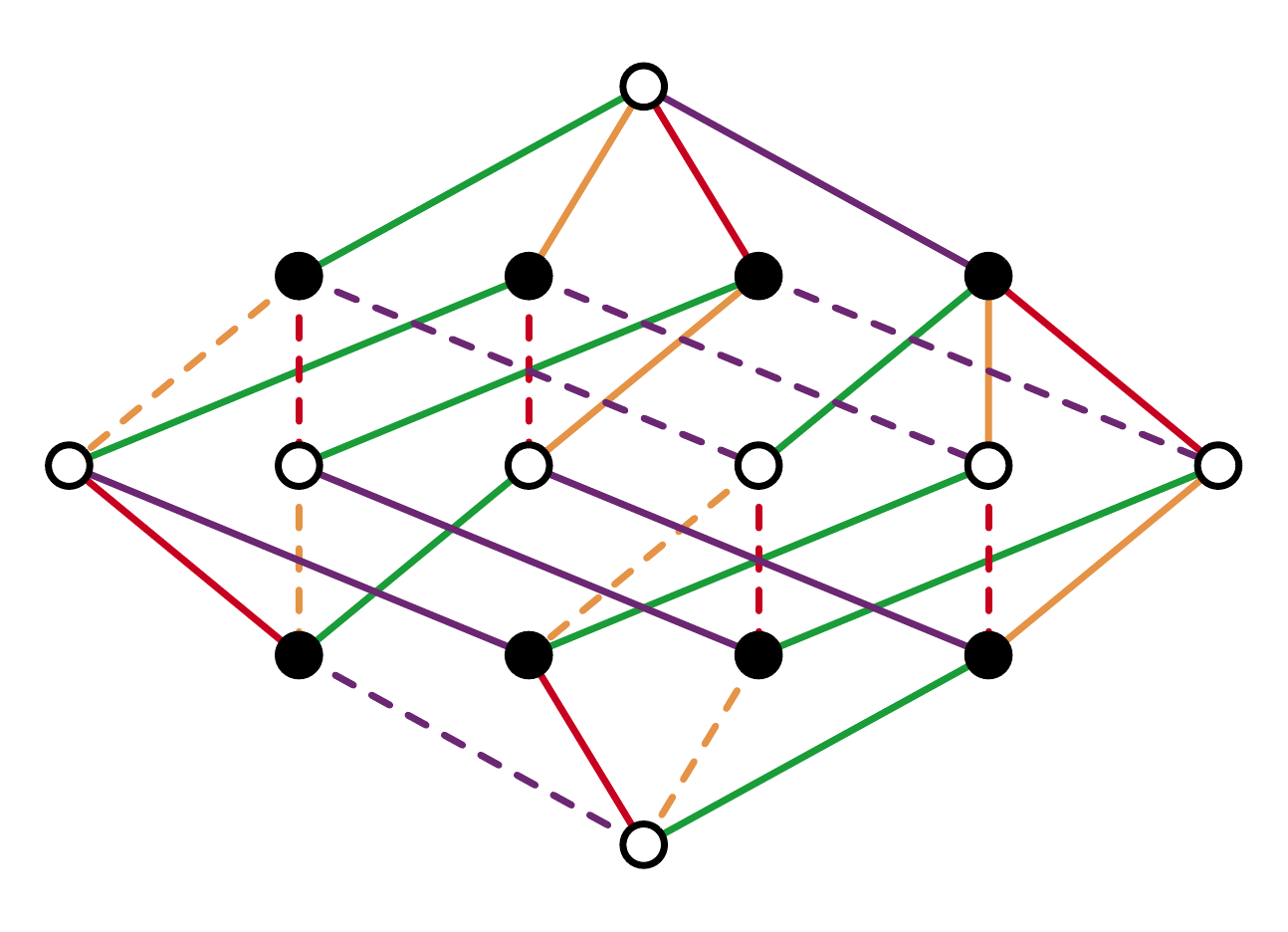}}  
\end{picture}
\vskip0.7in
\noindent
Figure 5: The minimum number of directed edges in the Garden baobab (left), and the associated Garden adinkra (right). All edges in the Garden adinkra are suppressed, but point from the lower node to the higher one.
\vskip0.1in
  \begin{picture}(-20,0)
  \put(-5,-110){ \includegraphics[width=0.45\columnwidth]{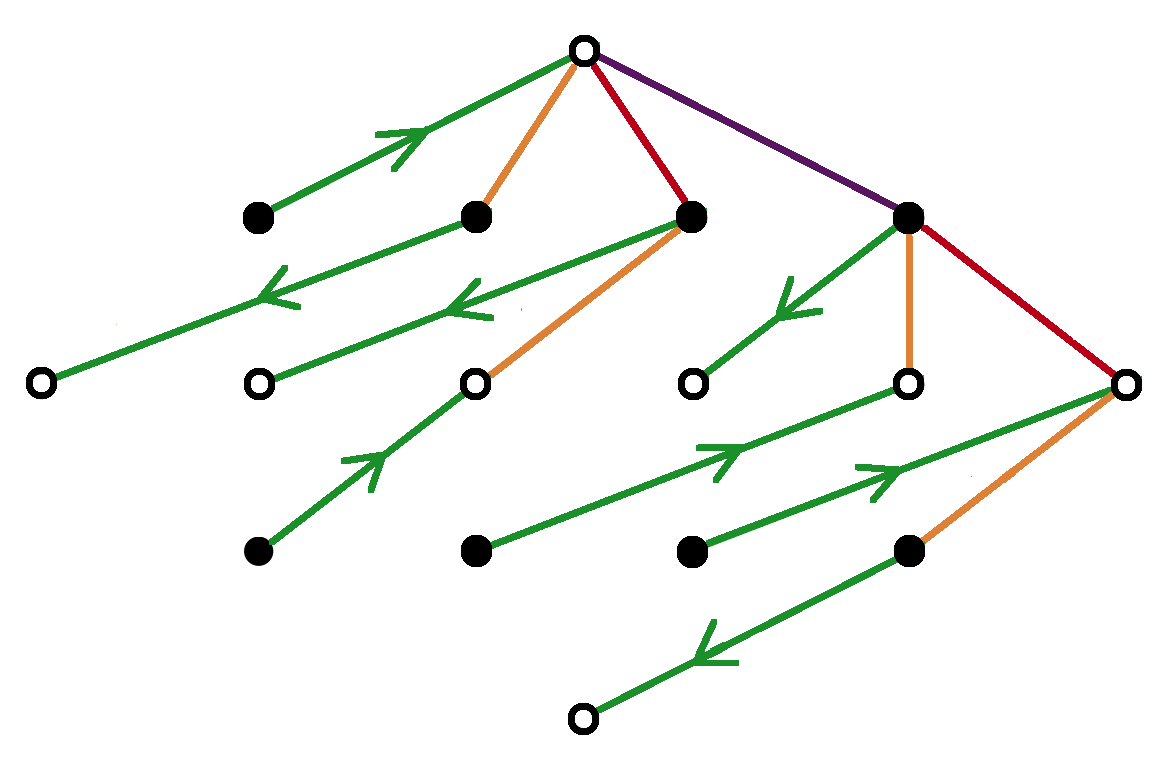}}
    \put(175,-85){ \includegraphics[width=0.43\columnwidth]{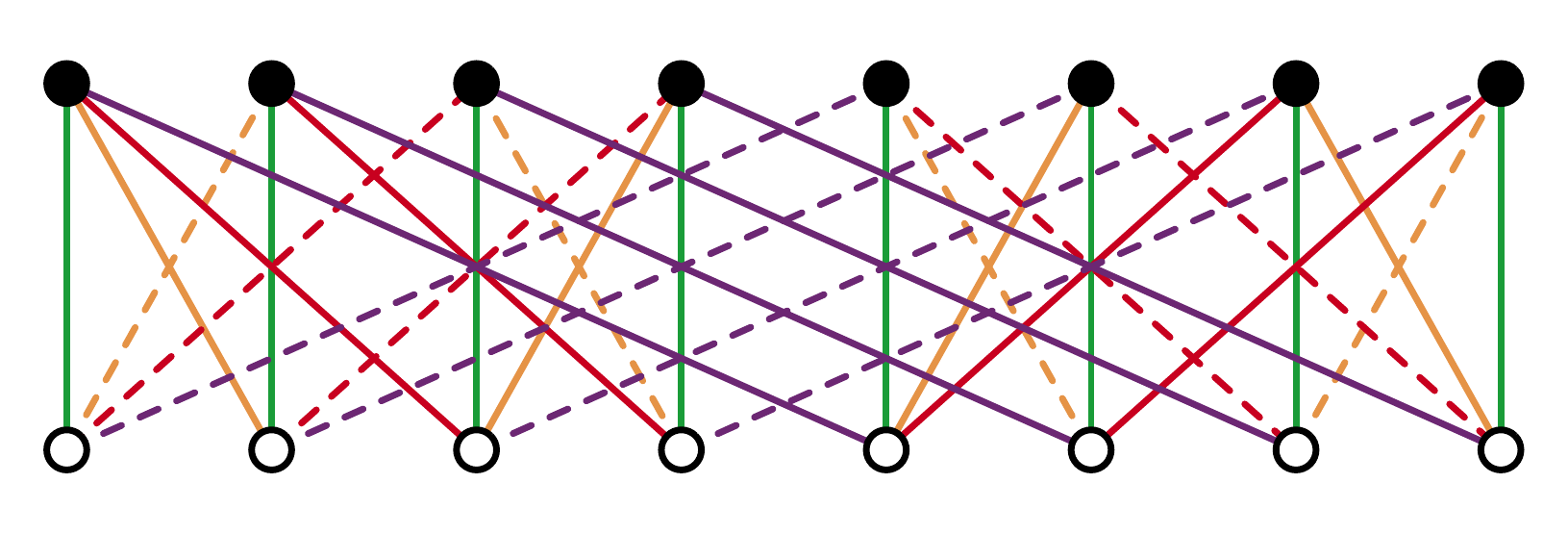}}  
\end{picture}
\vskip1.7in
\noindent
Figure 6: The maximum number of directed edges in the Garden baobab (left) needed to reproduce the Garden adinkra (right). As before, edge directions are suppressed but point from the lower node to the higher one.
\vskip0.2in

\noindent
By the Hanging Garden's theorem, a few pinned nodes acting as sources or sinks, determine all other edge directions \cite{YanThesis}. Therefore, the edges in a baobab are directed simply to determine what are sources and what are sinks and where they lie in the adinkra.

 It is, however, thus-far non-trivial to generally determine which edges must be directed in a baobab. It is obvious that directed edges must be attached to sources and sinks (which is always possible in a simply connected graph like a baobab). I conjecture, however, that all the other directed edges in the baobab, to use the terminology of previous publications, ``rank" sources and sinks by connecting at least one source to one sink \cite{YanThesis}. 

Given a basic understanding of adinkras and baobabs in Garden algebra, we can now look at how these graphs are generalized.

\vskip0.7in
\section{Adinkras For Other Algebras}

$~~~$I will now define adinkras and baobabs for more general Lie superalgebras.\newline

\noindent
{\bf Definition 0:}\newline
An {\it adinkra adjacency matrix} is an adjacency matrix for a single edge color of a weighted graph. If the graph is undirected, then the matrix is assumed symmetric. For a graph with arcs, each direction of an edge may represent a distinct value in the matrix, thus node $i$ directed to node $j$ will correspond to the matrix element $a^i_{j} = \alpha \neq a^j_{i}$. Furthermore, each edge may have a distinct weighting, thus multigraphs are isomorphic to a weighted simple graph possibly with loops.\newline

First, I will prove the following theorem.\newline

\noindent
{\bf Theorem 1:} Given a finite dimensional Lie superalgebra, $\mathfrak{L}$, with $m$ elements, one can construct a mixed weighted graph, $G(V,E)$, with $m$ edge colors that can graphically represent the elements.
 
\noindent {\bf Proof:}\newline 
 The arbitrary edge weighting and edge directions of $G(V,E)$ imply the adinkra adjacency matrix elements need not be correlated and can have any value in $\mathbb{R}$. Furthermore, every adjacency matrix is square, implying the adinkra adjacency matrix can be any $m \times m$ real matrix. Lastly, any number of edge colors can be added to the graph, to represent distinct matrices. By Kac's generalization of Ado's Theorem, $\mathfrak{L}$'s elements can similarly be represented by a set of $m \times m$ real matrices [6, 17], which by construction can be the adinkra adjacency matrices of some graph $G(V,E)$.\begin{flushright}$\Box$\end{flushright}

This leads to the next definition.\newline

\noindent
{\bf Definition 1:}\newline
A {\it Lie adinkra} is a graph with edge colors and weighted edges whose adinkra adjacency matrices represent the complete set of elements of a Lie superalgebra. Edges with the same color but distinct weights are visually distinguished, for example, by edge dashing. \newline

As with Garden adinkras, I define the Lie adinkra to have a chromotopology, therefore a Lie adinkra with strictly yellow, green, and red edges is not isomorphic to an adinkra with strictly blue, green, and orange edges. Finally, one can generalize Garden baobabs.

\noindent
{\bf Definition 2:}\newline
 The {\it Lie baobab} is a subgraph of a Lie adinkra that contains the least information necessary to reconstruct the adinkra knowing the Lie bracket values.\newline

The Lie baobab is always smaller than the Lie adinkra, because the Lie bracket values create a constraint to the edge weights of the adinkra. 

Lie adinkras and baobabs contain several nice properties that are useful when creating and comparing superalgebra representations.

As seen in Garden adinkras via the Hanging Gardens Theorem \cite{GraphTh}, Lie adinkras  can produce a representation as easily as representations can produce the graphs, suggesting a visually appealing way of constructing representations of finite dimensional superalgebras\footnote{For example, the elements ${\color{blue}i}$, ${\color{Green}j}$ and ${\color{red}k}$ are isomorphic to elements in {\bf su}(2), thus representations of {\bf su}(n) for $n\ge 3$ might be generalizable from the pattern seen here.}. The Lie baobab can distinguish representation isomorphism classes quickly because the minimal number of elements appear in each adinkra adjacency matrix. Matrix properties useful to determine the isomorphism class of a representation, can instead be modified for use with these sparser matrices. 

In addition, errors or erasures of edge dashing, direction, etc. in a Lie adinkra can be discovered and, in many cases, corrected, using the Lie superalgebra bracket values (and baobabs) as checks. This observation allows for the Lie adinkra to be encoded in binary format and bit errors corrected with forward error correction codes derived from the associated Lie superalgebra. In the next section, it will be shown that a message will be resilient to errors as long as a baobab is a subgraph of the resultant graph. If a bit is flipped, some baobabs will change but most will not, therefore, baobabs can act as agents in an error correcting ``democracy", by having a bit flipped back if the majority of the baobabs need for the bit to be flipped in order to make a consistent adinkra. 

Lastly, interpreting the graphs as error correction codes will allow for at least some graphs to be interpreted as a set of logic gates.

\vskip0.7in
\section{Logic Circuits And Baobabs}

$~~~$In this section I will explain some relationships between Lie adinkras, baobabs, and information theory. If a dashed edge is represented as a ``0" bit, and an undashed edge, a ``1" bit in a Garden adinkra, then the ``odd number" edge dashing property is equivalent to the function I call the ``not double exclusive or" (NDXOR)  gate: \newline

$~~~~~~~~~~~~~~~~~~~ \text{NDXOR}
:=\{\forall ~x,~y,~z~\in \mathbb{Z}_2|~\neg~(x~ \oplus~ y~ \oplus~ z)\}~~~~~~~~~~~~~~~~~~~~~~~~(10)$ \newline

\noindent
If 3 edges in a two-color cycle have an odd (even) number of dashed edges, the fourth edge must be undashed (dashed). The Garden baobab edge dashing can act as the ``input" bits, while the rest of the Garden adinkra's edge dashing can act as the ``output" bits. Edge directions have an equivalent gate,  $\neg(\text{NDXOR})$ or simply the DXOR gate, where  edges directed parallel to a trail around a two-color cycle can correspond to ``0" and edges directed anti-parallel to a trail, ``1"\footnote{There are some restrictions to the input, however. One cannot have a 111 or 000 input, as these would imply edges in a two-color cycle would all be directed head to tail, and thus there would be no source or sink, which is impossible for a Garden adinkra.}. Ignoring the edge directions for now, a Garden baobab can transform into its associated adinkra by hooking up NDXOR gates in cascade such that the dashing of 2-color 3-paths are the input to each gate.
For example, Fig. \# 7 shows how the edge dashing can be completed for a 3-color Garden adinkra, while Fig. \# 8 shows the equivalent action with binary gates.
\newline\newline\newline\newline\newline
  \vskip4in
  \begin{picture}(-20,0)
  \put(-20,-025){ \includegraphics[width=1\columnwidth]{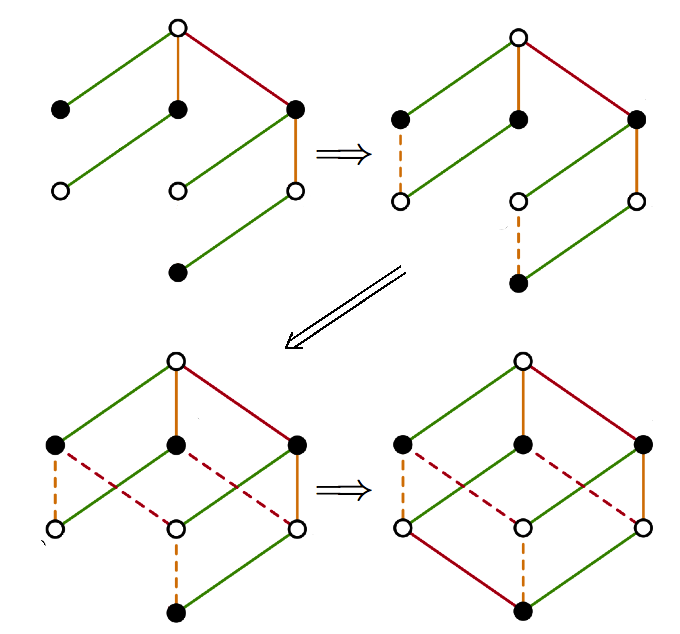}}
  \label{f:NewBaobab}
\end{picture}
\vskip.3in
\noindent
Figure 7: Construction of a Garden adinkra from a Garden baobab, based on the principle that all two-color cycles must have an odd number of dashed edges. Directed edges have been suppressed.
\vskip0.2in

Previous work relating to codes and Garden adinkras have focused on labeling nodes as codewords quotiented by a set of doubly even codes [5, 14, 15]. One could state that this paper simply generalizes this idea to include both edge dashing and edge directions, and is furthermore deeply related to these codes. For example, the set, $\{\mathcal{O}_i\}$, of a baobab has a 1-to-1 correspondence with a set of doubly even codes describing the Garden adinkra's chromotopology. This is because the codewords are meant to represent cycles of the form $\Gamma_i\Gamma_j\Gamma_k\Gamma_l...$ with $4m$ distinct edge colors, which can be the set $\mathcal{O}_p$ of some baobab cycle. In addition, the order of  $\mathcal{O}_p$ for a given cycle must be doubly even for the same reason as the code (see p.18-21 of \cite{YanThesis}). Because all $k$ cycles are linearly independent, these cycles are isomorphic to an $(n+k,k)$ doubly even code.

  \vskip1.3in
  \begin{picture}(-20,0)
  \put(-20,-15){ \includegraphics[width=1\columnwidth]{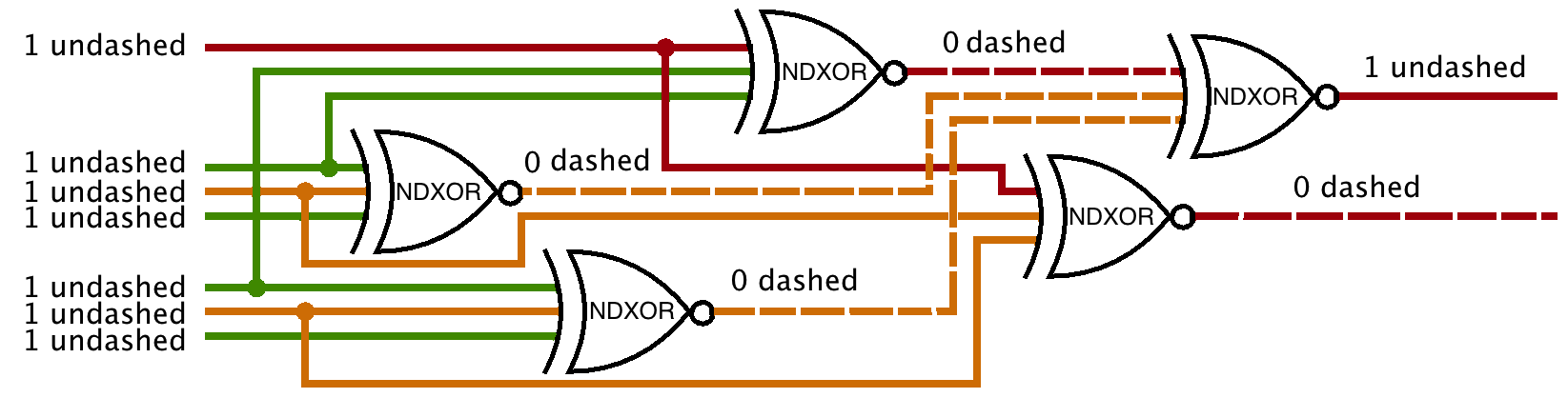}}
  \label{f:NewBaobab}
\end{picture}
\vskip0.3in
\noindent
Figure 8: The logic circuit corresponding to the construction of the Garden adinkra in Fig. \# 7.
\vskip0.2in

  Interestingly, Fig. \# 7 suggests that several edges can be removed or changed and the original adinkra could still be re-constructed. If edges are lost, they can trivially be recovered by using the Lie bracket values and a baobab sub-graph.
If an edge is flipped, however, the algebra will not be preserved throughout the graph. Figuring out which edges will allow for baobabs to all produce the same adinkra will say which edges were most likely changed.
  
This idea isn't unique to Garden algebra, however. There seems to be an entire class of new forward error correction block codes representing Lie superalgebras that can be adinkraizable. For example, any edge can be changed and corrected in the quaternion adinkra (see Fig. \# 9). If two edges change, it can still be detected, meaning the quaternion adinkra has a Hamming distance of 3. 
In general, messages can be written into a baobab's set of adinkra adjacency matrices, before they are ``covered up" by reconstructing the entire adinkra. Determination of errors simply amounts to checking whether the matrices follow their appropriate algebra\footnote{Although this could be done without the help of a baobab or an adinrka, both of these objects act as visual aids to writing a message.}.
\newline\newline
\newline\newline
\newline\newline
\newline\newline
\newline\newline
\newline\newline
\newline\newline
\newline\newline
\newline\newline
\vskip0.7in
  \begin{picture}(-20,0)
  \put(-20,-15){ \includegraphics[width=1\columnwidth]{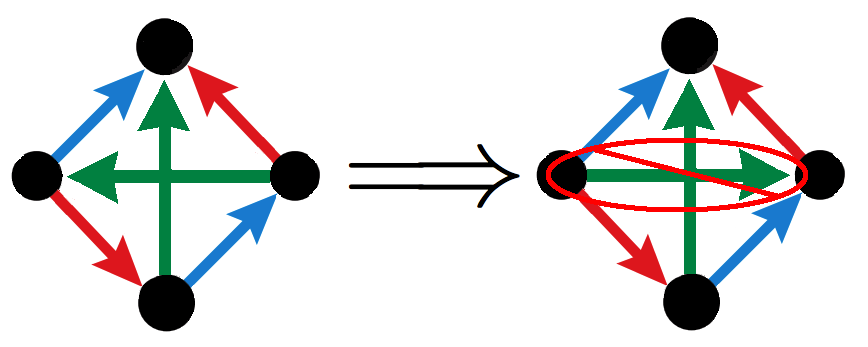}}
  \label{f:NewBaobab}
\end{picture}
\vskip0.3in
\noindent
Figure 9: Using the quaternion adinkra to detect and correct edge direction errors. If the horizontal green edge flips direction (right), it is immediately obvious because ${\color{blue} i}{\color{Green} j}{\color{red} k}\neq -1$, $\{{\color{blue} i},{\color{Green} j}\}\neq 0$, and $\{{\color{Green} j},{\color{red} k}\}\neq 0$. Furthermore, the only way to preserve quaternion algebra again (by flipping a single bit) is for the green edge to flip back.
 \vskip0.7in
 \section{Conclusion}
 
$~~~$ I have asked two important questions in this paper.

 Firstly, can one describe a Garden adinkras degrees of freedom? I show that one can with a new graph called a baobab. This graph is significantly smaller (in fact very nearly the smallest possible connected graph, the tree), and due to its simple topology, could be used to more easily distinguish non-isomorphic Garden adinkras. 
  
  Secondly, can adinkras and baobabs be generalized to describe other algebras? The answer is again yes, using the more general Lie adinkras, and Lie baobabs. 
  
  Surprisingly, these graphs can also represent a set of logic gates and forward error correction codes. When edges are changed from these graphs, they can still be detected as long as one baobab in the adinkra remains unaffected. 
  
  There is plenty of future work ahead, however. One important open problem is how to fully understand the relationship between the directed edges in a Garden adinkra and the corresponding edges in a Garden baobab. Without a complete understanding, it would be hard to determine how to add the directed edges to the baobab in general.
Furthermore, it is clear that a message can be ``covered up" via an adinkra, but if one forgets how the message was encoded, it seems to be extremely difficult to determine where the message was hidden in the graph. Therefore, it would be interesting to determine whether an adinkra can be an effective block cipher.

 \vskip1in
{\Large\bf Acknowledgments}

KB would like to thank Dr. S. James Gates of the University of Maryland as well as Dr. Tristan Hubsch of Howard University for their stimulating conversations. Lastly, the author is in gratitude to Kory Stiffler for his suggestions on how to improve the paper. Lie adinkras and Lie baobabs were drawn with the help of {\it Adinkramat} $\copyright$ 2008 by G. Landweber and \cite{AdnkGraph}.

\end{document}